\documentclass[twocolumn,nofootinbib]{revtex4-1}
\usepackage{graphicx}
\usepackage{epstopdf}
\usepackage{amsmath}

\newcommand{\bessi}{{I}}
\newcommand{\eq}{{\mathrm eq}}

\begin{document}
\title{Evolution of the moments of multiplicity distributions}
\author{Radka Sochorov\'a$^{a}$}
\author{Boris Tom\'a\v{s}ik$^{a,b}$}
\author{Marcus Bleicher$^{c,d,e}$}
\affiliation{$^a$\v{C}esk\'e vysok\'e u\v{c}en\'i technick\'e v Praze, 
FJFI, B\v{r}ehov\'a 7, 11519 Praha 1, Czech Republic}
\affiliation{$^b$Univerzita Mateja Bela, Tajovsk\'eho 40, 97401 Bansk\'a Bystrica, Slovakia}
\affiliation{$^c$Frankfurt Institute for Advanced Studies, 
Johann Wolfgang Goethe-Universit\"at, Ruth-Moufang-Strasse 1, 60438 Frankfurt am Main, Germany}
\affiliation{$^d$Institut f\"ur Theoretische Physik, 
Johann Wolfgang Goethe-Universit\"at, Max-von-Laue-Strasse 1, 60438 Frankfurt am Main, Germany}
\affiliation{$^e$GSI Helmholtz Center, Planckstr.~1, 64291 Darmstadt, Germany}


\begin{abstract}
Measured moments of the multiplicity distribution for a given sort of particles 
are used in the literature for the determination of the phase transition parameters 
of hot QCD matter in ultrarelativistic heavy-ion collisions. We argue that the 
subsequent cooling in the hadronic phase, however, may drive  the multiplicity distribution 
out of equilibrium. We use a master equation for the description of the 
evolution of the multiplicity distribution to demonstrate how the different moments depart away
from their equilibrium values. If such moments were 
measured and interpreted as if they were equilibrated, one would obtain different 
apparent temperatures from different moments. 
\end{abstract}
\maketitle


\section{Introduction}

Event-by-event
fluctuations of the identified particle number observed in relativistic heavy-ion  collisions  carry the promise to 
exactly locate  the state of the created QCD matter on the phase diagram at the time of hadron production
\cite{Asakawa:2000wh,Hatta:2003wn,Koch:2005vg,Ejiri:2005wq,Stokic:2008jh,Athanasiou:2010kw,Gavai:2010zn,Stephanov:2011pb,Gupta:2011wh}. 
The moments of the number distribution can be related to the susceptibilities 
which are expressed as derivatives of the partition function
\cite{Athanasiou:2010kw,Stephanov:2008qz}. 

The susceptibilities are usually related to a conserved quantum number, e.g.\ the 
baryon number or strangeness. 
A large variety of susceptibilities are currently determined by lattice QCD methods 
up to the fourth order \cite{Bazavov:2017tot}. 
On the experimental side, moments of the net proton number distribution are also
measured up to the fourth order \cite{Adamczyk:2013dal,Ohlson:2017wxu,Szala:2018rrr}. 
Unfortunately, baryon number cannot be measured, as neutrons are
not detected  in many current experimental set-ups. 
Nevertheless, there are arguments that claim that the protons are a good proxy for the 
baryon number \cite{Kitazawa:2011wh,Kitazawa:2012at}. 
In real collision events, the observed proton number fluctuations, however, are also influenced by conservation
laws \cite{Bzdak:2012an,Braun-Munzinger:2016yjz}.
In addition to net baryon number, the fluctuations of the number of strange particles are also measured
\cite{Adamczyk:2017wsl,Ohlson:2017wxu}.

From a comparison of experimental data to theoretical predictions of various  moments of the 
multiplicity distribution, temperature and chemical potentials of the created matter can be determined
\cite{Alba:2014eba,Bellwied:2018tkc,Bluhm:2018aei}. 
Note however, that the theoretical treatments---be it lattice QCD or the statistical model---use 
the grand-canonical formalism and assume equilibrium. The results obtained from such analyses
show some disagreement between temperatures obtained from fitting the first moments (i.e.~the yields)
\cite{Andronic:2017pug}
and those obtained from fitting the higher moments \cite{Alba:2014eba}. 

Our study is directly motivated by such a mismatch. We qualitatively show that in an ensemble of expanding and cooling 
fireballs different moments of the number distribution may acquire values that seemingly do not
correspond to each other if one tries to understand them with single temperature and chemical potential. 

To clarify this point, let us stress that after hadronisation inelastic collisions among hadrons still continue. 
Due 
to the decrease of the reaction rates they are unable to maintain the chemical composition so that it would 
fully respond to the changing temperature. 
In fact, the inelastic reactions alter the numbers of individual species 
and drive them off equilibrium.  Our treatment thus goes beyond \cite{Stephanov:2009ra} where no 
inelastic collisions after chemical freeze-out were assumed. 

Note further, that since we want to study the \emph{fluctuations} of multiplicities, we inherently 
study an ensemble of fireballs and the time evolution of multiplicity distribution across the ensemble. As
the distribution drops out of equilibrium, its moments may not be described by universal values of temperature and 
chemical potentials. This is the essence of the argument presented in this paper. 

The appropriate tool for studying the evolution of multiplicity distributions is a master equation. Generally, in contrast to rate 
equations,  master equations describe the evolution of the \emph{whole discrete probability distribution} and not just the evolution 
of the mean values. The description is related to the microscopic processes which can change multiplicity of the studied 
kind of particles. In Section~\ref{s:me} 
we use a specific master equation which also respects exact $U(1)$ charge conservation \cite{Ko:2000vp} and derive 
the equilibrium values for the first four moments. Then, in Section~\ref{s:therm} we look at how the formalism describes 
relaxation towards equilibrium. Phenomenologically relevant scenario of cooling is investigated in Section~\ref{s:cool}. 
Conclusions are presented  in 
Section~\ref{s:conc}.


\section{The master equation}
\label{s:me}

For our study we shall investigate multiplicity distributions  of species that conserve an abelian charge, e.g.\
strangeness, and undergo the reaction $a_1a_2 \leftrightarrow b_1 b_2$. Here, none of the involved species are identical 
to each other and it is understood that the $b$'s carry the conserved charge while the $a$'s do not. Also, it will 
be assumed that there is a sufficiently large pool of $a$'s which is basically untouched by this chemical process\footnote{%
For too small numbers of $a$'s, e.g.~in collisions at lower energies within the RHIC Beam Energy Scan programme, this 
assumption may not necessarily be warranted. This would lead to a modification of the master equation. We shall come to 
this point again in the discussion section.
}.
In \cite{Ko:2000vp} the master equation 
for such a process has been derived which describes the time evolution of the multiplicity distribution 
$P_n$ of species $b$. Here, $P_n$ is the probability to have $n$ pairs of those species. The master equation 
is formulated as 
\begin{multline}
\label{e:me}
\frac{dP_n}{dt} = \frac{G}{V} \langle N_{a_1}\rangle \langle N_{a_2} \rangle \left [ P_{n-1} - P_n \right ]\\
- \frac{L}{V} \left [ n^2 P_n - (n+1)^2 P_{n+1} \right ] \,  ,
\end{multline}
where $V$ is the effective volume and $G$, $L$ stand for the momentum-averaged cross section 
of the gain process ($a_1a_2 \to b_1 b_2$) and the loss process ($b_1b_2 \to a_1 a_2$), respectively
\begin{eqnarray*}
G & = & \langle \sigma_G v \rangle \label{e:gain}
\\
L & = & \langle \sigma_L v \rangle\,  .
\end{eqnarray*}
We suppressed writing out explicitly that the $P_n$'s are functions of time. 

Equation (\ref{e:me}) can be solved in order to obtain the evolution of all $P_n$'s. 
If $G$, $L$, and $V$ are constant, then it must describe the approach towards equilibrium.
This is best studied, if the master equation is put 
into dimensionless form
\begin{equation}
\frac{dP_n}{d\tau} = \varepsilon \left [ P_{n-1} - P_n \right ] - \left [ n^2 P_n - (n+1)^2 P_{n+1} \right ]\,  ,
\label{e:metl}
\end{equation}
by scaling the time with relaxation time $\tau_0$
\begin{equation}
\tau = \frac{t}{\tau_0}\, , 
\end{equation}
where 
\begin{eqnarray}
\label{e:tau0}
\tau_0 & = & V/L \\ 
\varepsilon & = & \frac{G}{L} \langle N_{a_1}\rangle \langle N_{a_2} \rangle\,  .
\end{eqnarray}
The equilibrium distribution can be then derived  with the help of the generating function
\cite{Ko:2000vp}
\begin{equation}
g(x,\tau) = \sum_{n=0}^\infty x^n P_n(\tau)\,  ,
\end{equation}
where $x$ is an auxiliary variable. 

The generating function is instrumental in calculating the factorial moments of the multiplicity 
distribution. It obeys the normalisation condition 
\begin{equation}
g(1,\tau) = \sum_{n=0}^\infty P_n(\tau) = 1\, ,
\label{e:norm}
\end{equation}
and gives
\begin{subequations}
\label{e:moms}
\begin{eqnarray}
g'(1,\tau) & = & \sum_{n=0}^\infty n P_n(\tau) = \langle n \rangle \\
g''(1,\tau) & = & \sum_{n=0}^\infty n (n-1) P_n(\tau) = \left \langle n (n-1) \right \rangle \\
g^{(3)} (1,\tau) & = & \sum_{n=0}^\infty n (n-1) (n-2) P_n(\tau) \nonumber\\
&& = \left \langle \frac{n!}{(n-3)!}\right \rangle \\
g^{(4)} (1,\tau) & = & \sum_{n=0}^\infty n (n-1) (n-2)(n-3) P_n(\tau)
\nonumber \\ &&  = \left \langle \frac{n!}{(n-4)!}\right \rangle\,  . 
\end{eqnarray}
\end{subequations}

In order to find the equilibrium distribution, one derives from the  
master equation (\ref{e:metl}) the equation for the time evolution of $g(x,\tau)$ \cite{Ko:2000vp}
\begin{equation}
\frac{\partial g(x,\tau)}{\partial \tau} = 
(1-x) \left (
x \frac{\partial^2 g}{\partial x^2} + \frac{\partial g}{\partial x} - \varepsilon g 
\right )\, . 
\end{equation}
The equilibrium solution is found if the right-hand-side is set equal to 0 \cite{Jeon:2001ka}
\begin{equation}
x \frac{\partial^2 g}{\partial x^2} + \frac{\partial g}{\partial x} - \varepsilon g = 0
\end{equation}
It reads
\begin{equation}
g_0(x) = \frac{\bessi_0(2\sqrt{\varepsilon x})}{\bessi_0(2\sqrt{\varepsilon})}
\end{equation}
which fulfils the normalisation condition (\ref{e:norm}). Here, $\bessi_0(x)$ is the modified Bessel function.

The equilibrium distribution is then 
\begin{equation}
P_{n,eq} = \frac{\varepsilon^n}{\bessi_0(2\sqrt{\varepsilon}) (n!)^2}\, .
\end{equation}

Through equations (\ref{e:moms}) the factorial moments can be calculated.\footnote{%
The first and second factorial moments have been  calculated 
in \cite{Ko:2000vp,Jeon:2001ka}.}
\begin{subequations}
\label{e:evls}
\begin{eqnarray}
\langle n \rangle_{\eq}  & = & \sqrt{\varepsilon} 
\frac{\bessi_1(2\sqrt{\varepsilon})}{\bessi_0(2\sqrt{\varepsilon})} \\
\langle n (n-1) \rangle_{\eq}  & = & - \frac{1}{2} \sqrt{\varepsilon}
\frac{\bessi_1(2\sqrt{\varepsilon})}{\bessi_0(2\sqrt{\varepsilon})} \nonumber \\
& & + \frac{1}{2}\varepsilon 
\frac{\bessi_2(2\sqrt{\varepsilon}) + \bessi_0(2\sqrt{\varepsilon})}{\bessi_1(2\sqrt{\varepsilon})} \\
\left \langle \frac{n!}{(n-3)!}\right \rangle_{eq} & = & 
\frac{3}{4} \sqrt{\varepsilon}\frac{\bessi_1(2\sqrt{\varepsilon})}{\bessi_0(2\sqrt{\varepsilon})} \nonumber \\
&&- \frac{3}{4} \varepsilon\left ( 1 +  \frac{ \bessi_2(2\sqrt{\varepsilon})}{\bessi_0(2\sqrt{\varepsilon})}\right )\nonumber \\
&&+ \frac{1}{4} \varepsilon^{3/2} 
\frac{\bessi_3(2\sqrt{\varepsilon}) + 3\bessi_1(2\sqrt{\varepsilon})}{\bessi_0(2\sqrt{\varepsilon})}\\
\left \langle \frac{n!}{(n-4)!}\right \rangle_{eq} & = & 
-\frac{15}{8} \sqrt{\varepsilon} \frac{\bessi_1(2\sqrt{\varepsilon})}{\bessi_0(2\sqrt{\varepsilon})} \nonumber\\
&& + \frac{15}{8} \varepsilon \left (\frac{\bessi_2(2\sqrt{\varepsilon})}{\bessi_0(2\sqrt{\varepsilon})} +1\right )
\nonumber \\
&&-\frac{3}{4} \varepsilon^{3/2} 
\frac{3\bessi_1(2\sqrt{\varepsilon}) +\bessi_3(2\sqrt{\varepsilon})}{\bessi_0(2\sqrt{\varepsilon})} 
\nonumber \\
&& + \frac{1}{8} \varepsilon^2 \left ( 3 + 
\frac{4\bessi_2(2\sqrt{\varepsilon}) + \bessi_4(2\sqrt{\varepsilon})}{\bessi_0(2\sqrt{\varepsilon})}\right )
\nonumber \\ && 
\end{eqnarray}
\end{subequations}

These analytical expressions for  equilibrium values of the factorial moments allow us to assess separately 
the departure from equilibrium for different orders.  

Other characteristics of the distribution, like the central moments, cumulants, skewness, or kurtosis etc., can 
be calculated by combinations of these factorial moments. 

To scale out the total number of particles, we shall also study the scaled factorial moments
\begin{subequations}
\begin{eqnarray}
F_2 & = & \frac{\langle n (n-1) \rangle}{\langle n \rangle^2}\\
F_3 & = & \frac{\left \langle\frac{n!}{(n-3)!}\right \rangle}{\langle n \rangle^3}\\
F_4 & = & \frac{\left \langle\frac{n!}{(n-4)!}\right \rangle}{\langle n \rangle^4} \,   .
\end{eqnarray}
\end{subequations}


\section{Thermalisation}
\label{s:therm}

The goal is to study why and 
how different moments of multiplicity distribution may indicate different temperatures. 
We therefore first study how the various moments relax towards their equilibrium values in 
an environment with constant temperature. 

If the temperature and cross-section are fixed, then the only time scale in the problem does not 
change and we can use eq.~(\ref{e:metl}) for the evolution of the multiplicity distribution. 

The presented results have been obtained from a simulation with binomial initial conditions
\begin{subequations}
\label{e:binini}
\begin{eqnarray}
P_0(\tau = 0) & = & 1-n_0\\
P_1(\tau = 0) & = & n_0\\
P_i(\tau = 0) & = & 0\, \qquad \mbox{for}\,\, i>1\, ,
\end{eqnarray}
\end{subequations}
where 
\[
n_0 = \langle n \rangle (\tau = 0)
\]
is the mean multiplicity of species $b$ at the beginning.

\begin{figure}
\begin{center}
\includegraphics[width=0.47\textwidth]{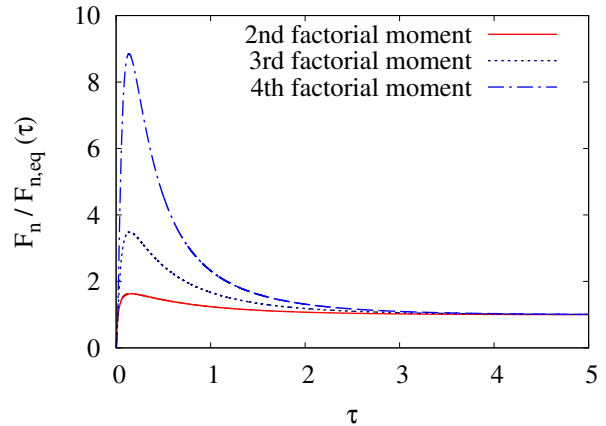}
\end{center}
\caption{Relaxation of the scaled factorial moments as functions of dimensionless time $\tau$.
Binomial initial conditions with $\varepsilon = 0.1$ and $n_0 = 0.005$ have been used.}
\label{f:relax}
\end{figure}

The evolution of second to fourth scaled factorial moments divided by their equilibrium values 
(from eqs.~(\ref{e:evls})) is shown in Figure~\ref{f:relax}.
%

%
The value of the parameter $\varepsilon$ has been set to 0.1 and the initial mean to $n_0 = 0.005$. 
Note that we have obtained qualitatively similar results also with other sets of parameters. Initial conditions which follow the 
Poisson distribution lead to similar rates of relaxation although the initial part of the dependence is 
different. 

The higher moments are more sensitive to the 
shape of the distribution function beyond just its width.
Thus one might have anticipated,
that they  more easily depart  from their equilibrium values and it might take longer time for them to 
reach equilibrium. 
While the first statement holds,  the second assertion is not fulfilled.
Higher 
moments indeed depart further away from the equilibrium values in a non-equilibrium situation. 
However, they actually relax in the same time as the lower moments.


\section{Non-equilibrium cooling}
\label{s:cool}

The fireballs produced in ultrarelativistic heavy-ion collisions cool down rapidly. It is therefore expected that  
the number distribution departs from equilibrium. In our simulation we shall assume 
that  the system is  equilibrated at the hadronization temperature $T=165$~MeV. 
Due to subsequent expansion, the temperature decreases quickly. 
For a system that stays in equilibrium a 
lower temperature would correspond to a different 
multiplicity distribution. However, a prerequisite to maintain equilibrium is to 
ensure that the creation and annihilation reactions are 
capable of running at rates comparable to the expansion rate, otherwise the equilibrium will be lost. We
explore what this means for the values of the moments. 

To set up the environment, we shall assume Bjorken one-dimensional boost invariant expansion, where 
the proper volume grows linearly
\begin{equation}
V(t) = V_0 \frac{t}{t_0}
\label{e:vexp}
\end{equation}
and the temperature drops according to 
\begin{equation}
T^3(t) = T_0^3\frac{t_0}{t}\, .
\end{equation}
In the calculations we have set $V_0 = 125\, \mbox{fm}^3$. The temperature will start at the 
value of 165~MeV and drop down to 100~MeV. The latter is typically the temperature of the kinetic
freeze out\footnote{%
There is no general agreement in the literature concerning the kinetic freeze out temperature. While 
fits with the blast-wave model without resonances yield   temperatures around 120~MeV for collisions 
at RHIC \cite{Adamczyk:2017iwn}, fits with resonance decays included give temperatures of 
100~MeV \cite{Melo-prep} and lower \cite{Rode:2018hlj}, depending on the details of the model.
}. 
Motivated by the femtoscopic measurements 
we set the final time to 10~fm/$c$.  This then leads to $t_0 = 2.2~\mbox{fm}/c$

For the actual calculation we must also choose the particular inelastic process. Processes with 
too small cross-sections will not be able to change anything on the multiplicity distributions while 
those with very large cross-sections will practically simultaneously adjust them to the decreasing 
temperature. The relevant time scale is the relaxation time, given in eq.~(\ref{e:tau0}). The interesting 
regime is when the relaxation time is comparable to the inverse expansion rate and the lifetime of the fireball. 

We consider the reaction  
$\pi^+n \leftrightarrow K^+ \Lambda$. 
For the moment we shall use a parametrisation of the cross-section \cite{Cugnon:1984pm}
\begin{equation}
\label{e:xc}
\sigma_{\pi N}^{\Lambda K} = \left \{ 
\begin{array}{lc}
0\, \mbox{fm}^2 & \sqrt{s} < \sqrt{s_0}\\
\frac{0.090 (\sqrt{s} - \sqrt{s_0})}{0.091}\, \mbox{fm}^2 &  \sqrt{s_0}\le\sqrt{s}<\sqrt{s_0}+0.09\,\mbox{GeV}\\
\frac{0.0090}{\sqrt{s} - \sqrt{s_0}} \, \mbox{fm}^2 & \sqrt{s} \ge \sqrt{s_0}+0.09\,\mbox{GeV}
\end{array}
\right .
\end{equation}
where $\sqrt{s_0}$ is the threshold energy of the reaction and the energies are given in GeV.
Unfortunately, the flux-averaged gain and loss term with this cross-section are too small 
and lead to too low reaction rates. In order to proceed with qualitative studies, 
the cross-section has been scaled up by hand so that the 
relaxation time is a few fm/$c$. Note that the gain term (eq.~(\ref{e:gain})) of this reaction 
is small due to the rather high threshold, which is about 530~MeV above the masses of the incoming particles, 
while the temperature is  
lower than 165~MeV.  This indicates that the reaction rate might increase considerably if the 
threshold is lowered, for example through the decrease of the hyperon mass in baryonic matter.
This possibility will be investigated below. 

\begin{figure}
\begin{center}
\includegraphics[width=0.47\textwidth]{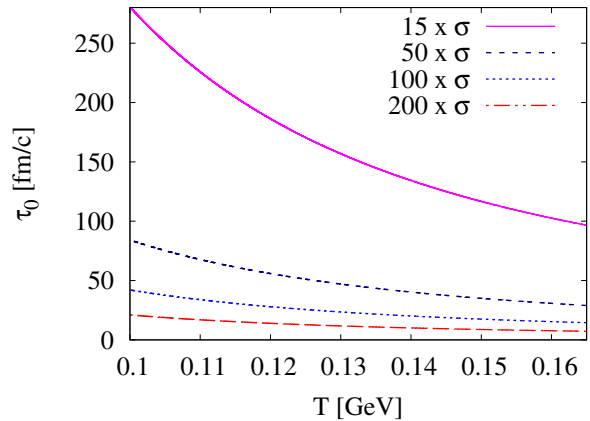}
\end{center}
\caption{
Dependence of the relaxation time on temperature for scaled cross-sections. There are 15 pions and 
10 neutrons and the initial volume of  125~fm$^3$ expands according to eq.~(\ref{e:vexp}).
}
\label{f:trelax}
\end{figure}

In Fig.~\ref{f:trelax} we show how  the relaxation time changes as the fireball exapands and cools down. 
%
%
The relaxation time decreases with increasing temperature and/or with increasing 
cross-section. 
We have performed calculations with all scales of the cross-sections indicated in Fig.~\ref{f:trelax}.

\begin{figure}
\begin{center}
\includegraphics[width=0.47\textwidth]{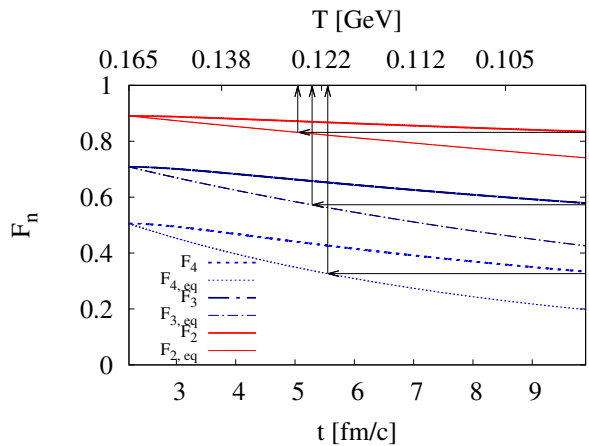}
\end{center}
\caption{
Evolution of the scaled factorial moments. Thick lines: values calculated through the master equation (\ref{e:me}), 
thin lines: equilibrium values calculated from relations (\ref{e:evls}). 
}
\label{f:facmom}
\end{figure}

First, we present in Fig.\ref{f:facmom} the evolution of the scaled factorial moments.
%
%
Due to decreasing temperature the moments change, but the reaction rate is too low to keep them in equilibrium. 
As observed above, the relative departure from equilibrium is larger for moments of higher order. 

We can now demonstrate the potential danger in case of extraction of the freeze-out temperature 
from the different moments. Suppose that the system breaks up at final (kinetic) temperature of 100~MeV, i.e.\
the moments assume their final values at this point. How does one usually 
infer the temperature from such a measurement? 
One \emph{assumes thermalisation}. The moments of a thermalised system would have evolved along the 
thin curves in Fig.~\ref{f:facmom}. So the assumption of  thermalisation means that one assumes that the 
thin lines provide the correct description of the moments. However, in reality the moments evolved along 
the thick lines in Fig.~\ref{f:facmom}.
The arrows in that Figure
show how the apparent freeze-out temperature would be extracted. The actual observed final value of a thick line is 
projected horizontally on the corresponding thin line (Fig.~\ref{f:facmom}) and the apparent 
temperature is read off from the projected point. 
We can see that such a procedure can  lead to different values of apparent 
temperature if different moments are 
used.

Factorial moments are usually not measured. More common are the central moments
\begin{subequations}
\begin{eqnarray}
\mu_1 & = & \langle n \rangle = M\\
\mu_2 & = & \langle n^2 \rangle - \langle n\rangle^2 = \sigma^2\\
\mu_3 & = & \langle (n-\langle n\rangle )^3\rangle \\
\mu_4 & = & \langle (n-\langle n\rangle )^4\rangle\, ,
\end{eqnarray}
\end{subequations}
or  their derived characteristics: the skewness
\begin{equation}
S = \frac{\mu_3}{\mu_2^{3/2}} \,  ,
\end{equation}
and the kurtosis
\begin{equation}
\kappa = \frac{\mu_4}{\mu_2^2} - 3\,  .
\end{equation}
Their equilibrium values can all be calculated from proper combinations of the factorial moments and using 
the equilibrium values derived in eqs.~(\ref{e:evls}).

\begin{figure}
\centerline{\includegraphics[width=0.49\textwidth]{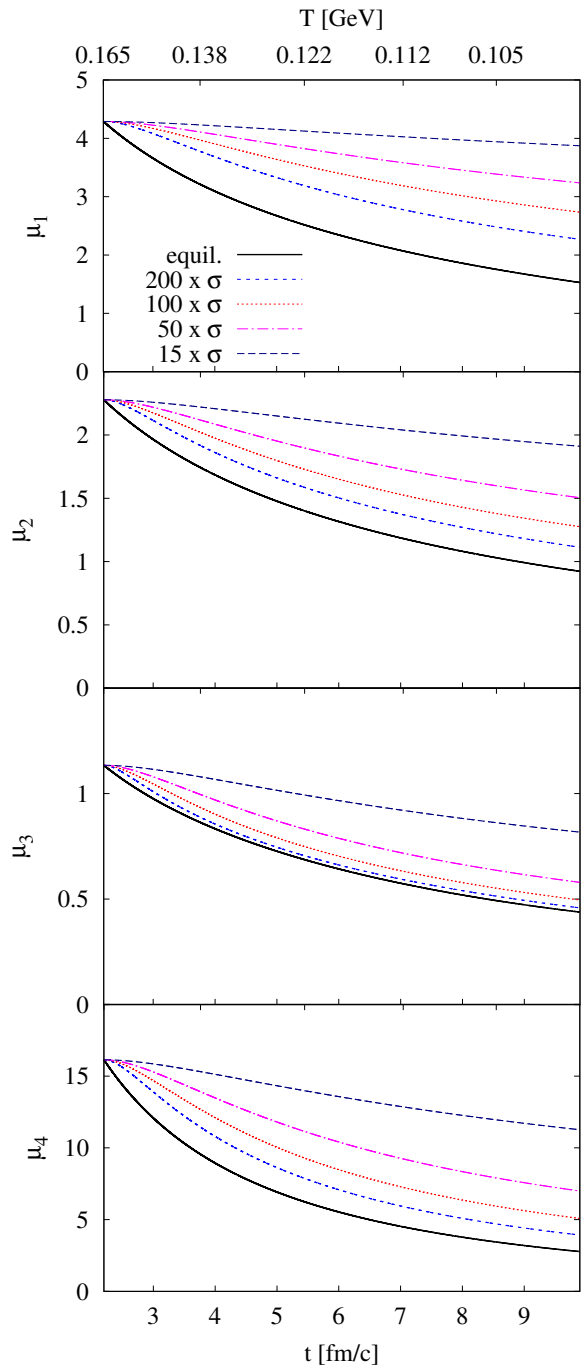}}
\caption{ 
Evolution of the first four 
central moments (from top to bottom). Different curves on  the same panel show results for different cross-sections. 
Solid lines show the equilibrium values. 
}
\label{f:central}
\end{figure}

We plot the evolution of central moments for different scales of the cross-section in Fig.~\ref{f:central}.
%
%
As expected, larger cross-sections maintain the calculated values closer to the equilibrium ones. 
Also, moments of different orders generally indicate different apparent freeze-out temperatures, 
if interpreted as equilibrium values. 

\begin{figure}
\centerline{\includegraphics[width=0.49\textwidth]{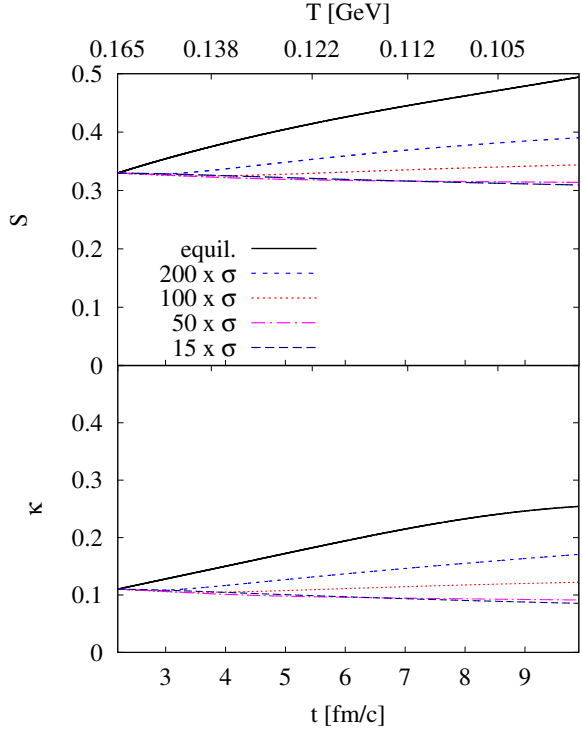}}
\caption{
Evolution of the skewness (upper panel) and the kurtosis (lower panel). 
Different curves on  the same panel show results for different cross-sections. 
Solid lines show the equilibrium values.  
}
\label{f:Sk}
\end{figure}

The skewness and kurtosis are even more interesting (Fig.~\ref{f:Sk}). 
%
%
Their equilibrium values grow as the temperature decreases. This is not true, however, for  the numerically calculated 
curves. Only those with larger cross-sections do increase, while those with smaller cross-sections decrease. 
Clearly, the apparent freeze-out temperature could only be determined from those numerically calculated 
curves, which increase with the temperature.

\begin{figure}
\centerline{\includegraphics[width=0.49\textwidth]{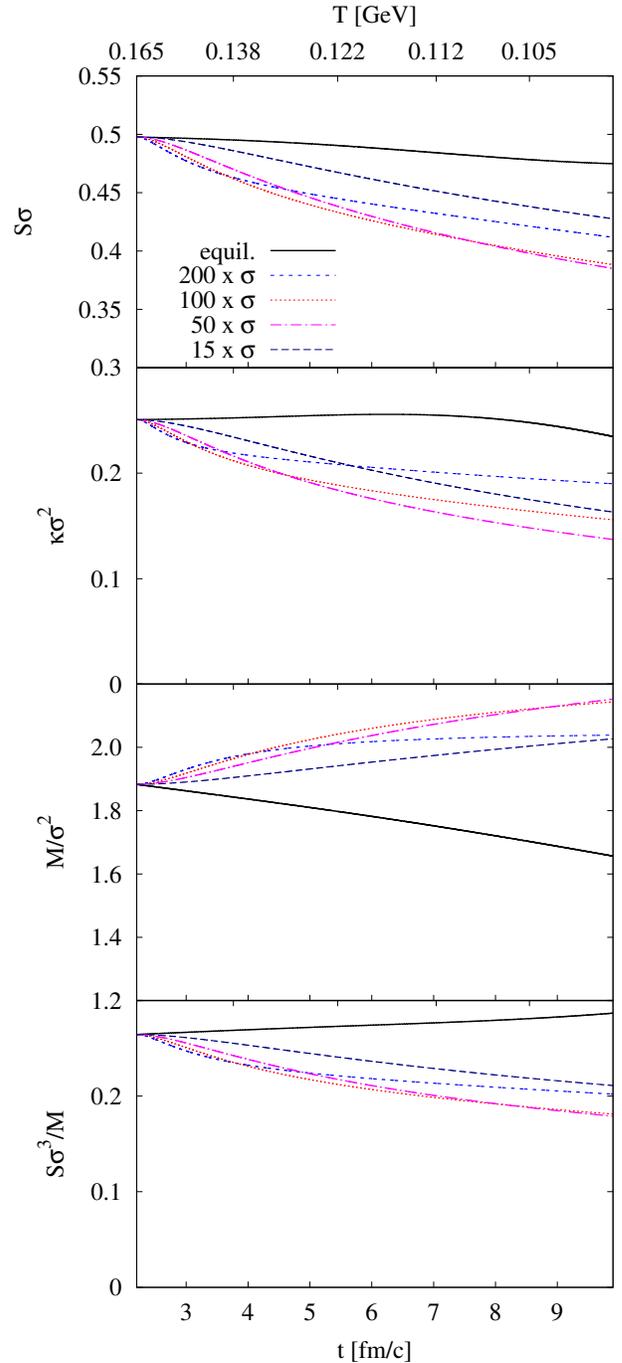}}
\caption{
Evolution of the volume-independent ratios. 
Different curves on  the same panel show results for different cross-sections. 
Solid lines show the equilibrium values. 
}
\label{f:Rs}
\end{figure}

For the sake of completeness, we also look at the volume-independent ratios which are often 
measured due to their easier comparison with theory. These are
\begin{subequations}
\begin{eqnarray}
R_{32} & = & \frac{\mu_3}{\mu_2} = S\sigma\\
R_{42} & = & \frac{\mu_4}{\mu_2} - 3\mu_2 = \kappa\sigma^2\\
R_{12} & = & \frac{\mu_1}{\mu_2} = \frac{M}{\sigma^2}\\
R_{31} & = & \frac{\mu_3}{\mu_1} = \frac{S\sigma^3}{M}\,  .
\end{eqnarray}
\end{subequations}
The evolution of these ratios for different cross-sections, together with the equilibrium values 
at actual temperatures, are presented in Fig.~\ref{f:Rs}. 
%
%
We find that some of the ratios, notably $R_{12}$ and $R_{31}$, evolve qualitatively differently from the equilibrium 
value for any value of the cross-section. 

In summary, simple factorial and central moments behave so that for 
larger cross-section we see how they approach the equilibrium behaviour. However, when they 
are combined into more complicated observables, like skewness, kurtosis, or the 
volume-independent ratios, the departure from the equilibrium is considerable and even 
qualitative. We conclude that these observables are actually very fragile with respect to 
non-equilibrium chemical evolution and very easily depart from values which can be 
interpreted in terms of equilibrium distribution. 

The previous studies presented in this work have been performed with cross-sections that were scaled unrealistically 
high for the given reaction. The aim was to use this reaction as a proxy for any other processes which can produce
the $b$-species. This is acceptable, because our conclusions from 
the study are only qualitative.

Nevertheless, it is also unrealistic to assume that the masses and cross-sections in hot and dense baryonic matter 
keep their vacuum values. Moreover, by lowering the mass of the hyperon,
also the threshold for the reaction is lowered, and its rate may grow due to the increase of 
the available phase space. We shall assume rather extreme and simplified dependence 
of the hyperon mass on baryon density 
\begin{equation}
m_\Lambda(\rho_B) = \frac{\rho_0 - \rho_B}{\rho_0} m_{\Lambda 0} + 
\frac{\rho_B}{\rho_0} m_{p}
\label{e:lmass}
\end{equation}
so that the hyperon mass becomes identical to that of the proton at the highest baryon density $\rho_0$ at 
which our calculation starts, and returns to the vacuum value 
$m_{\Lambda 0}$ if baryon density vanishes. The cross-section 
in eq.~(\ref{e:xc}) is modified by replacing the threshold $\sqrt{s_0}$ by the new value $m_K+m_\Lambda(\rho_B)$.

\begin{figure*}
\centerline{\includegraphics[width=0.9\textwidth]{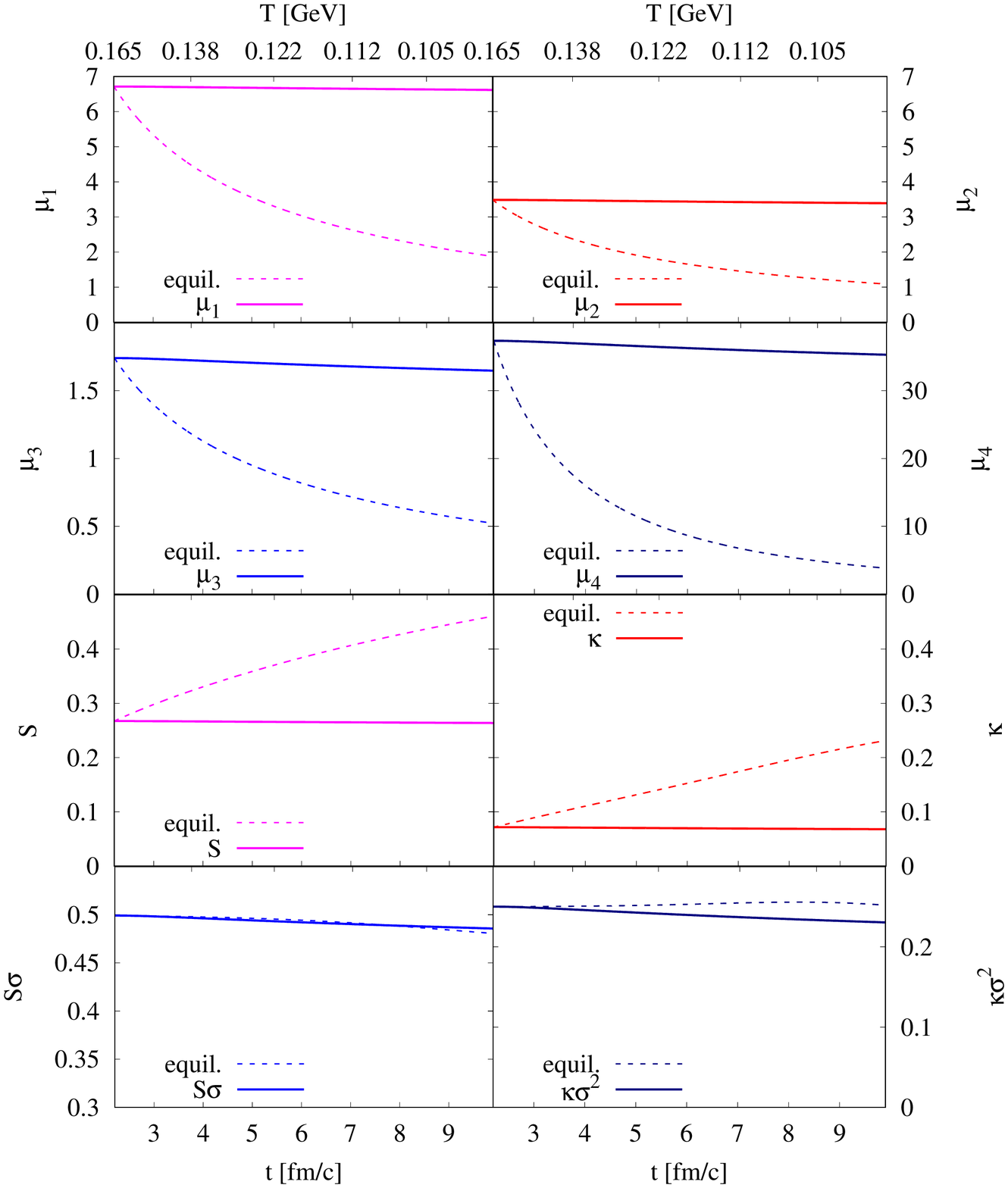}}
\caption{
Central moments, skewness, kurtosis, and volume-independent ratios $S\sigma$ and 
$\kappa\sigma^2$ for the scenario with density-dependent mass of $\Lambda$. 
Thick solid lines: numerically calculated evolution; thin dotted lines:
equilibrium values at the given temperature.
}
\label{f:massdrop}
\end{figure*}

Selected results from the scenario with density-dependent mass of the hyperon are plotted in 
Figure~\ref{f:massdrop}.
%
%
The rate of the reaction is clearly too small to keep the system in chemical equilibrium. However, thanks to 
the increase of the cross-section at the highest baryon densities the system is clearly driven away 
from the state which is present at the initial temperature of 165~MeV. All central moments 
are slightly decreasing, because the multiplicity of strange particles goes down. The hierarchy of the departure 
from the initial value is such that it grows with the order of the moment.
Only slight changes are seen for the skewness and the kurtosis and somewhat stronger departure 
is observed for the volume independent ratios $S\sigma$ and $\kappa\sigma^2$. Interestingly,
for the latter even the equilibrium values seem not to change much, although this is rather accidental.

Nevertheless, in realistic fireballs there are other channels that can change the numbers of kaons and/or 
lambdas and so we have to expect a variation of moments yet larger than indicated by these 
calculations.  
We conclude that with a realistic scenario the evolution of the composition of the fireball may be 
capable of influencing the fluctuations of the  particle number distribution to a measurable extent.



\section{Discussion}
\label{s:conc}

In this work, we have explored  one of the effects  that may influence the higher moments of the multiplicity
distribution of identified particles. In general, if hadronisation is followed by a re-scattering phase, the multiplicity distribution may 
become off-equilibrium. This means that not only the average number of particles may change, but also the higher 
moments  depart from their equilibrium values. 

The master equation (\ref{e:me}) adopted from \cite{Ko:2000vp} and used here is suitable for the 
description of single reaction channel which produces a $U(1)$ conserved charge. It can be used for 
the studies of rare species, like charm or bottom. Of course, the measurement of moments of their 
multiplicity distribution is a challenge which is very hard to overcome. 

We have applied the formalism to one reaction channel which produces strangeness,
and we have studied the fluctuations of strange particles. The higher moments of kaon multiplicity 
distribution have been measured within the RHIC Beam Energy Scan program by the STAR collaboration 
\cite{Adamczyk:2017wsl}. We have made an attempt to apply our calculations to the interpretation 
of those data. However, this actually revealed that our calculation is only a part of a
more complex description. 
Firstly, there are certainly other channels that influence the number distribution of strange particles. 
Secondly---and probably more importantly---we do not know the  initial conditions for the 
evolution. 

Let us also come to the point raised in the introduction of the master equation: what would happen if 
we cannot replace the numbers of $a$-species just by their averages. The master equation would slightly 
change, but the general feature of our results remains: the different moments of the number distribution 
depart from their equilibrium values, and higher moments do that more than the lower moments.

As a side project we have also looked at the isospin-randomising reactions which turn protons into 
neutrons and vice versa. Such reactions have large cross-sections and no threshold. Hence, 
they are very frequent. Consequently, we observed that the moments of multiplicity distribution of protons 
do not change even with decreasing temperature if they are started in equilibrium. This is in line 
with the findings of \cite{Kitazawa:2011wh,Kitazawa:2012at} and it is a good news for the measurements
of proton number fluctuations, which try to relate the measured moments to the baryon number susceptibilities of the 
matter at the point of  hadronization. 

It appears as an interesting question, whether the formalism of master equations can be also used for the 
description of the deconfined phase of the collision. Recall that the evolution is interesting if the reaction rates 
are comparable to the rate of temperature decrease. This rules out the description of light quarks, which are 
produced and destroyed too easily. It also disqualifies the description of charm and bottom, which are too heavy 
to be produced in the quark-gluon plasma. What remains is the production of strange quarks, which might be 
interesting in a regime where QGP is produced, yet strangeness is not chemically equilibrated. 
The treatment, however might not 
be easy. Firstly, a different master equation must be derived which takes into account production from $q\bar q$
annihilations as well as $gg$ reactions. Secondly, it is expected that in this window of collision energies the system 
may spent an important portion of its time at temperatures just above the hadronisation. There, it is strongly coupled, 
interactions are non-perturbative, and the microscopic description with quarks and gluons becomes complicated 
\cite{Peshier:2005pp}.

Coming back to hadronic reactions investigated in this paper, 
we conclude that inelastic reactions in a system with decreasing temperature may   alter 
the individual moments of the multiplicity distribution differently. Importantly, they can push the moments 
away from their values at the 
beginning of cooling. Such moments are being measured with the hope that their precise values will help 
us to better pinpoint the position of the strongly interacting matter on the phase diagram. Therefore,
a word of caution must be raised that the values of the moments at hadronisation may be severely 
altered in subsequent evolution. Master equations are effective tools for the investigation of these
effects.


\begin{acknowledgments}
We thank E.E.~Kolomeitsev,  L.~Laff\'ers, and M.~\v{S}umbera for enlightening discussions. 
This work was supported by the grant 17-04505S of the Czech Science Foundation (GA\v{C}R).
This work was supported by collaboration grant in framework of the German-Slovak PPP programme and 
the COST Action CA15213 THOR. 
This work was supported by HIC for FAIR within the Hessian LOEWE initiative.
BT acknowledges support by VEGA 1/0348/18 (Slovakia). 
\end{acknowledgments}

\end{document}